\documentstyle[aps,prb,twocolumn,graphics]{revtex}

\begin{document}

\twocolumn[\hsize\textwidth\columnwidth\hsize\csname@twocolumnfalse\endcsname 
\title {\ Disorder Effects in  Superconducting Multiple Loop Quantum Interferometers}
\author{ Ch. H\"{a}ussler, J. Oppenl\"{a}nder, and N. Schopohl}
\address{Institut f\"{u}r Theoretische Physik, Eberhard-Karls-Universit\"{a}t T\"{u}bingen,
Auf der Morgenstelle 14, 72076 T\"{u}bingen, Germany}

\maketitle

\begin{abstract}
A theoretical study is presented on a number $N$ of resistively shunted Josephson junctions
connected in $\it parallel$ as a disordered $1D$ array by superconducting wiring in such a 
manner that there are $N-1$ individual SQUID loops with arbitrary shape formed.
Under a constant current bias $ I>I_{c}$, and irrespective of the 
degree of the disorder, but depending on the strength of magnetic field $ B$,
all junctions in the array oscillate at the $\it same$ frequency $\nu_{B}$. 
Computer simulations of the full non linear dynamics of a disordered junction array reveal:
(i) the frequency $\nu_{B}$ is $\it not$ a periodic function of $B$,
(ii) in the overdamped junction regime $\nu_{B}$ displays
a sharp $\it global$ minimum around  $B=0$. For zero inductive coupling the problem 
becomes equivalent to a  $\it virtual$ single junction model.
\end{abstract}
\bigskip
]

So-called weak links, or Josephson junctions, are the basic active elements
of superconductor quantum electronics. A key feature of a weak link between
two superconductors, $1$ and $2$, is the property that there can flow a 
{\em dissipationless} macroscopic supercurrent $I_{s}\left( \varphi \right)$ 
due to the tunneling of Cooper pairs\cite{Josephson} with charge $2e$. This
supercurrent depends on the gauge invariant phase difference 
$
\varphi =\Theta _{1}-\Theta _{2}+\frac{2e}{\hbar c}\int_{1}^{2}<d{\bf s},\,%
{\bf A>}
$
of the macroscopic BCS pairing wavefunctions on either side of the weak
link. Josephson junctions made with modern fabrication techniques
\cite{Hypres} often have a sandwich type layered geometry, with a thin non
superconducting tunneling barrier in the middle between two thick
superconducting electrodes. In recent time also other types of weak links,
for example of the bi-crystal type, became important in high-temperature
superconductors\cite{Gross,Koelle}. For an ideal {\it S-I-S} junction
the supercurrent is connected to the phase difference $\varphi$ across the
tunneling barrier by 
$
I_{s}\left( \varphi \right) =I_{c}\sin \varphi.
$
It is important to realize that the supercurrent $I_{s}$ flows stationary
provided it does not exceed a characteristic critical current $I_{c}$, the
so-called Josephson critical current, which determines the maximum
dissipationless current that can flow across a tunneling barrier. In
general, $I_{c}$ depends on the material properties of the junction, on
temperature $T$, and also on magnetic field ${\bf B=\mathop{\rm rot}A}$.
Applying to a Josephson junction a bias current $I$ with a {\em constant}
strength $I>$ $I_{c}$, there appears a rapidly oscillating
voltage signal $V\left( t\right)$ across the junction, which determines the
rate of change of the time dependent phase difference $\varphi \left(t\right)$ 
according to 
\begin{equation}
\hbar \,\partial _{t}\varphi \left( t\right) =2eV\left( t\right)
\label{2nd Josephson}
\end{equation}
This is the fundamental non stationary Josephson relation which governs the
physics of weak superconductivity\cite{Josephson}. So, for $I>$ $I_{c}$
there flows, besides the dissipationless supercurrent $I_{s}$, also a
dissipative normal current $I_{n}$ in the junction, whose physical origin is
the transfer of single (unpaired) electrons.

Within the range of validity of the {\it RCSJ} model\cite{Likharev}, the
dissipative current may be described with sufficient accuracy as a
superposition of an ohmic current, characterized by a parallel ohmic shunt
resistance $R$, and a displacement current, which is characterized by a
parallel geometric shunting capacitance $C$ describing electric polarization
inside the tunnelling barrier. The total junction current $I$ is then: 
\begin{equation}
I=I_{c}\sin \varphi \left( t\right) +\frac{\hbar }{2eR}\,\partial_{t}
\varphi \left( t\right) +\frac{\hbar C}{2e}\,\partial _{t}^{2}\varphi
\left( t\right)  \label{junction current}
\end{equation}
The time average 
\begin{equation}
\left\langle V\right\rangle =\lim\limits_{t\rightarrow \infty }\frac{1}{t}
\int_{0}^{t}dt^{\prime }V(t^{\prime })=\frac{\hbar }{2e}\lim\limits_{t\rightarrow \infty}
\frac{\varphi \left( t\right)-\varphi \left(0\right) }{t}  \label{time average}
\end{equation}
is the {\it dc} voltage part of the in general not sinusoidal voltage signal 
$V(t)$ across the electrodes of a Josephson junction. For a strongly
overdamped junction, $C=0$, one finds, assuming a {\em constant} bias
current $I>I_{c}$, a relatively simple formula\cite{Likharev}: 
$
\left\langle V\right\rangle =R\,\sqrt{I^{2}-I_{c}^{2}}
$.
The {\it dc} voltage $\left\langle V\right\rangle$ is connected to the
oscillation frequency $\omega =2\pi \nu$ of the voltage signal $V\left(t\right) $ by: 
\begin{equation}
h\,\nu =2e\,\left\langle V\right\rangle
\end{equation}
This result for the voltage response function $\left\langle V\right\rangle$
of a weak link suggests a spectroscopic interpretation. When a Cooper pair
is transferred from the superconducting side $1$ to the superconducting side 
$2$ of the junction, under conditions where $I>I_{c}$, a microwave photon
with energy $2e\left\langle V\right\rangle $ is released in the form of one
quantum of electromagnetic radiation\ (Josephson radiation\cite{Yanson}).

As far as macroscopic quantum interference is concerned, it was actually
known \cite{London} long before the discovery of the Josephson effects, that magnetic flux
threading the area of a superconducting ring, made out of a material that is
thick compared to the magnetic penetration depth $\lambda$, should be
quantized in units of the flux quantum $\Phi _{0}=\frac{hc}{2e}$.

Technical applications of the physics of weak superconductivity\cite
{Likharev,SQUIDs} include ultrasensitive quantum interferometers,
which indeed combine the afore mentioned Josephson effects with flux
quantization. Consider, as indicated schematically in Fig.(1a), a standard
two junction {\it SQUID} (for simplicity with symmetric junction parameters)
under the {\it dc} current bias $I>$ $2I_{c}$. Such a device is actually a
flux-to-voltage transducer\cite{Likharev}. Let $\Phi =\left\langle {\bf B},\,
{\bf a}\right\rangle =$ $\left| {\bf B}\right| \left| {\bf a}\right| \cos
\alpha $ be the magnetic flux threading the orientated area element ${\bf a}$
of the superconducting {\it SQUID} loop, where $\alpha $ is the angle
between the normal vector of the orientated area element and the magnetic
field vector ${\bf B}$, as depicted schematically in Fig.(1a). The total
magnetic field, ${\bf B=B}^{\left( 1\right) }{\bf +B}^{\left( 2\right) }$,
is then a superposition of the {\em primary } external magnetic field 
${\bf B}^{\left( 1\right) }$, which generates the flux 
$\Phi ^{\left( 1\right)}=\left\langle {\bf B}^{\left( 1\right) },\,{\bf a}\right\rangle $ 
one wants to detect, and a {\em secondary } magnetic field ${\bf B}^{\left( 2\right) }$
that results, for example, from the inductance $L$ (or other impedance
effects) in the circuit. The screening current $I_{sc}$ circulating in the 
{\it SQUID }loop leads to a total flux 
$\Phi =\Phi ^{\left( 1\right)}\,+\Phi ^{\left( 2\right) }$. 
Dependent on the secondary flux term, $\Phi^{\left( 2\right) }=-L\cdot \,I_{sc}$, 
there exists an optimal size 
$\left| {\bf a}_{L}\right| $ for any {\it SQUID} loop \cite{SQUIDs}. 
A dimensionless measure for the inductance of such a loop is 
$\beta _{L}=\frac{L\cdot I_{c}}{\Phi _{0}}$. 
Note also that a {\em two} junction {\it SQUID} cannot be directly
employed as a detector of {\em absolute }strength of external magnetic
field. This is because the voltage response function 
$\left\langle V_{xy}\right\rangle $ of the {\it SQUID}, i.e. the time average of the
rapidly oscillating voltage signal $V_{xy}\left( t\right) $ across the nodes 
$x$ and $y$ of the circuit, is a {\em periodic} function of the strength of
external magnetic field, see Fig.(1a).

A straightforward extension of the standard two junction {\it SQUID }is
sketched in Fig.(1b). This is a $1D$ array of $N$ adjacent Josephson
junctions \cite{Likharev} connected in {\em parallel}. In particular, the
area elements of the $N-1$ {\it SQUID }loops formed in this manner are all
equal, e.g. ${\bf a}_{n}={\bf a}_{L}$ for all $n$. The voltage response
signal $\left\langle V_{xy}\right\rangle $ vs. strength 
$\left| {\bf B}^{\left( 1\right) }\right| $ of external magnetic field of this array has
the same period than a standard two junction {\it SQUID} with loop area
$\left| {\bf a}_{L}\right| $, see Fig.(1b).

A more general quantum interference device is obtained when the area
elements ${\bf a}_{n}$ of the $N-1$ loops in the array differ in size and,
possibly, in orientation, as depicted schematically in Fig.(1c).
If the sizes $\left| {\bf a}_{n}\right| $ of the orientated area elements ${\bf a}_{n}$ 
of the individual superconducting loops are chosen in a {\em disordered} 
fashion the voltage response function $\left\langle V_{xy}\right\rangle$
vs. $\left| {\bf B}^{\left( 1\right) }\right| $ becomes {\em nonperiodic},
see Fig.(1c). Taking into account inductive couplings among the currents in 
the circuit, the maximum loop size in the disordered array should coincide 
with the corresponding optimal loop size of a standard two junction SQUID,
i.e. $\max\left| {\bf a}_{n}\right| =\left| {\bf a}_{L}\right| $. 
The voltage response signal $\left\langle V_{xy}\right\rangle$ 
vs. strength of magnetic field of a {\it disordered} junction array is,
under a suitable {\it dc} current bias $I$, indeed a {\em unique} function
of $\left| {\bf B}\right| $ around its narrow {\it global} minimum at $\left| {\bf B}\right| =0$. 
This suggests that it should be possible, e.g. by
measuring control current(s) flowing through the wires of a set of suitably
orientated compensation coil(s), to reconstruct {\it absolute} strength,
orientation and even the phase of an incident primary magnetic field signal,
i.e. to determine the full vector ${\bf B}^{\left( 1\right) }(t)$.

The $n$-th Josephson junction in the array has, within the range of
validity of the {\it RCSJ} model, optional individual junction parameters 
$R_{n}$, $C_{\,n}$ and $I_{c,\,n}$. The corresponding current $I_{n}$
flowing through the $n$-th Josephson junction is, according to Eq.(\ref{junction current}), 
determined by the gauge invariant phase difference 
$\varphi _{n}\left( t\right) $ across that junction. The total current $I$
flowing through the nodes $x$ 
\begin{figure}
\begin{tabular}{c}
\\[-0.6cm]
\resizebox*{77 mm}{!}{\includegraphics{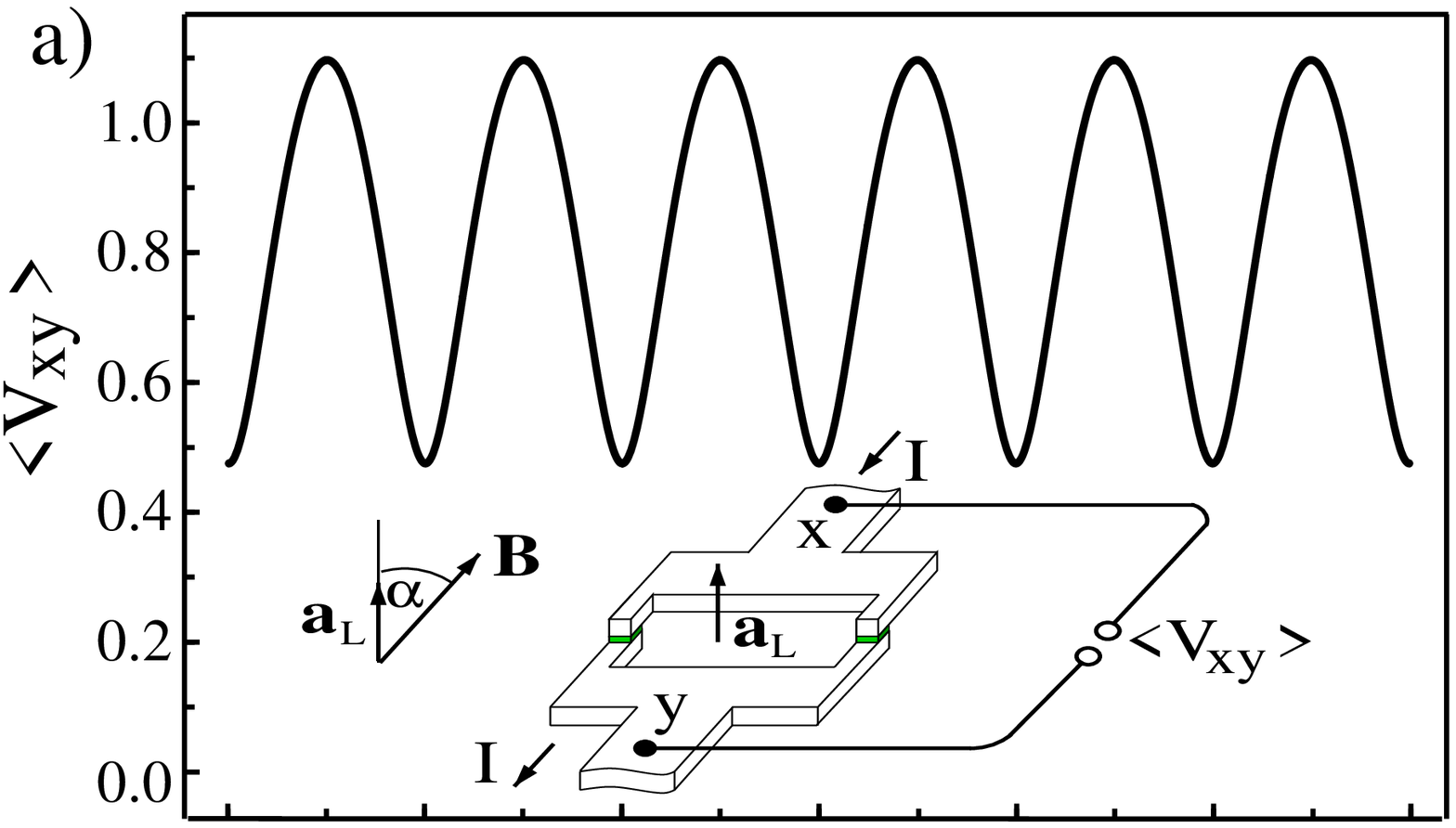}}\\[-0.6cm]
\resizebox*{77 mm}{!}{\includegraphics{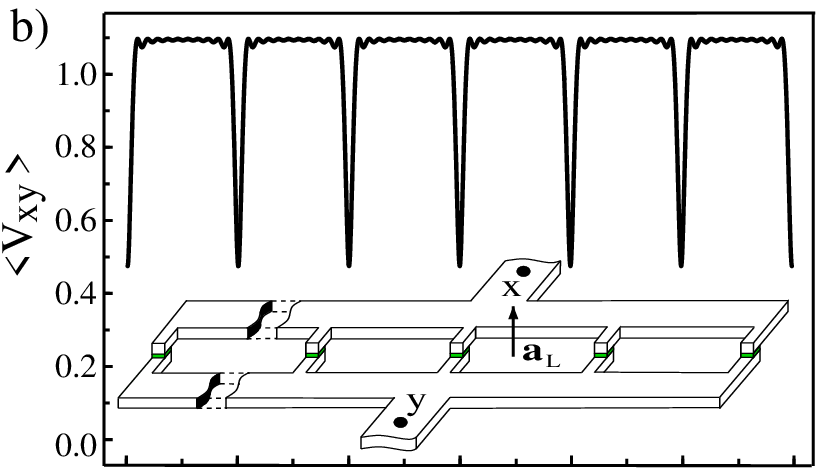}}\\[-0.6cm]
\resizebox*{77 mm}{!}{\includegraphics{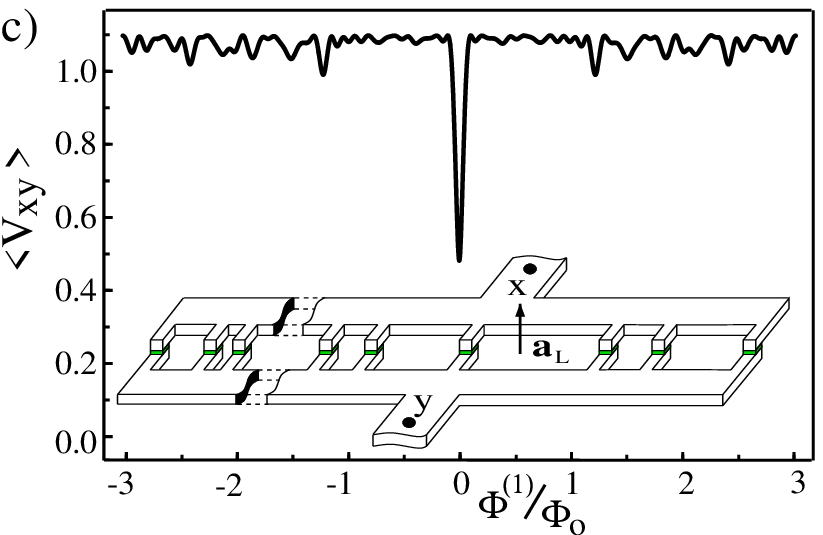}}
\end{tabular}
\caption{Voltage response $\left\langle V_{xy}\right\rangle$ in units of  
$I_c \, R$  vs. external flux $\Phi ^{\left( 1\right)}$ through largest area 
element ${\bf a}_{L}$ of interferometer for bias current $I=1.1\,N I_c$ and inductance
$\beta_L=0$: 
a) symmetrical SQUID $(N=2)$, b) periodic 1D array $(N=11)$, c) disordered 1D array (N=18), 
but with same total area as b).}
\end{figure}
\noindent
and $y$, respectively, of the circuit is then
obtained from Kirchhoff's rule as the {\em phase sensitive} superposition of
the individual junction currents $I_{n}$: 
\begin{eqnarray}
\label{Kirchoff}
I=\sum_{n=0}^{N-1}\left[ I_{c,\,n}\sin \varphi _{n}(t)+(\frac{\hbar C_{\,n}}{2e}\,
\partial_{t}^{2}+\frac{\hbar }{2eR_{\,n}}\partial _{t}\;) \;\varphi _{n}\left( t\right) \right]
\end{eqnarray}
Note that the gauge invariant phase differences $\varphi _{n}$ 
of adjacent Josephson junctions in the array are not independent, 
but are connected to each other by the condition of flux quantization: 
\begin{equation}
\varphi _{n}-\varphi _{n-1}=\frac{2\pi }{\Phi _{0}}\,\left\langle {\bf B\,},
{\bf a}_{n}\right\rangle \mathop{\rm mod} 2\pi  \label{flux quantization}
\end{equation}
Here $\left| {\bf a}_{n}\right| $ is the area of the superconducting loop
connecting adjacent Josephson junctions numbered as $n$ and $n-1$,
respectively, and ${\bf B}$ denotes the magnetic field threading the
orientated area element ${\bf a}_{n}$ of this loop. Note that Eq.(\ref{flux quantization}) 
applies quite generally, provided the superconducting
material, out of which the connecting loops are made, is thick compared to
the magnetic penetration depth $\lambda $. In this case there exists a path
inside the wire connecting, say, junction $n$ with its neighbor junction $n-1 $, 
on which the superfluid velocity field ${\bf v}_{s}$ becomes
negligible small. So, $\hbar {\bf \nabla }\Theta =\frac{2e}{c}{\bf A}$ along
this path. Since all junctions in the array are connected in parallel, the rapidly
oscillating voltage $V_{n}(t)$ at the electrodes of a particular Josephson
junction, numbered as $n\ $in the array, is related to the signal $V_{xy}(t)$
between the nodes $x$ and $y$ of the circuit by 
\begin{equation}
V_{xy}(t)=V_{n}(t)+\int_{{\small x\rightarrow n}\rightarrow {\small y}}
\left\langle d{\bf s},\ {\bf E}\left( t\right) \right\rangle  \label{V_AB}
\end{equation}
By Faraday's law the electric field ${\bf E}$ along an integration path $x\rightarrow n\rightarrow y$, 
that starts at node $x$, traverses the
tunneling barrier of the $n$-th Josephson junction just once, and then
terminates at node $y$, is directly connected to the time derivative of the
flux threading the area elements of the {\it 1D} array. Once the signal 
$V_{0}(t)$ $=$ $\frac{\hbar }{2e}\partial_{t}\varphi_{0}(t)$ is known the
other voltage signals $V_{n}(t)$ across the electrodes of the $n$-th
junction follow from
\begin{equation}
V_{n}(t)-V_{n-1}(t)=\frac{1}{c}\partial _{t}\,\left\langle {\bf B}\left(t\right) 
{\bf \,},{\bf a}_{n}\right\rangle
\end{equation}

Taking into account the Biot-Savart type inductive couplings\cite{Häußler et al.} 
among the currents flowing in the circuit prohibits further
simplification. However, it follows directly from Eq.(\ref{flux quantization}) 
that one can indeed eliminate from Eq.(\ref{Kirchoff}) all phase variables 
$\varphi _{n}(t)$ in favor of a single phase, for example $\varphi _{0}(t)$,
provided the extremely simplifying assumption is made, that all inductive
couplings vanish. In this case the problem of $N$ coupled Josephson
junctions is mapped onto a virtual {\em single} Josephson junction model.
With $\frac{1}{R}=\frac{1}{N}\sum_{n=0}^{N-1}\frac{1}{R_{\,n}}$, 
 $T_{N}=\frac{\hbar }{2e}\frac{1}{I_{c}\,R}$ and $\phi \left( t \right)=\varphi _{0}(t)$ 
there results a {\em scalar} differential equation determining the phase difference $\phi \left( t \right)$:
\begin{eqnarray}
&&\left| S_{N}\left( {\bf B}\right) \right| \, \sin \left[ \phi \left(t \right) 
+\delta _{N}\left( {\bf B}\right) \right] + T_N \left(RC\,
\partial_t^{2}+\partial _t\right) \phi \left(t\right)
\\
&&= J_{N}-\frac{2\pi }{\Phi _{0}} \, T_N \left(RC\, \partial _t^{2}\,
\left\langle {\bf B}\left(t \right) \,,{\bf a}_{C} \,\right\rangle
+\,\partial_t \,\left\langle {\bf B}\left(t \right) \,,{\bf a}_{R}\,
\right\rangle \right) \nonumber
\end{eqnarray}
where we have defined (${\bf a}_{0}={\bf 0}$)
\begin{eqnarray}
&S&_{N}({\bf B})=\frac{1}{N}\sum_{n=0}^{N-1}\frac{I_{c,\,n}}{I_{c}}\exp \left[ 
\frac{2\pi i}{\Phi _{0}}\sum_{m=0}^{n}\left\langle {\bf B}\,,\,{\bf a}_{m}
\right\rangle \right]  \label{structure factor}\\
&{\bf a}&_{R}=\frac{1}{N}\sum_{n=0}^{N-1}\frac{R}{R_{\,n}}\sum_{m=0}^{n}\,{\bf a}_{m}\;,\;\;
{\bf a}_{C}=\frac{1}{N}\sum_{n=0}^{N-1}\frac{C_{\,n}}{C}
\sum_{m=0}^{n}\,{\bf a}_{m} \nonumber 
\end{eqnarray}
and $J_{N}=\frac{I}{N\,I_{c}}$, $I_{c} =\frac{1}{N}\sum_{n=0}^{N-1}I_{c,\,n}$,
$C=\frac{1}{N}\sum_{n=0}^{N-1}C_{n}$.
The complex function $S_{N}\left({\bf B}\right) =\left|
S_{N}\left( {\bf B}\right) \right| e^{i\delta _{N}\left( {\bf B}\right) }$
denotes the characteristic {\em structure factor} of the {\it 1D} Josephson
junction array, as defined in Eq.(\ref{structure factor}). 
It is an extremely responsive function of strength and orientation of
magnetic field, and it is strongly affected by the choice of the individual
area elements ${\bf a}_{m}$. In general $\left| S_{N}\left( {\bf B}\right)
\right| $ is very sensitive to permutations among the ${\bf a}_{m}$'s.

In the overdamped junction regime, $C=0$, under conditions
where a constant current $I$ is biased such that 
$1\geq \left| S_{N}\left({\bf B}\right) \right| \,/\,J_{N}\equiv $ $\sin \alpha _{B}$, and assuming
for simplicity a homogeneous static magnetic field ${\bf B}$ (and also time
independent area elements ${\bf a}_{m}$), one finds an exact solution for
the phase difference $\varphi _{0}(t)$: 
\[
V_{0}(t)=\frac{\hbar }{2e}\partial _{t}\varphi _{0}(t)=I_{c}R
\frac{J_{N}^{2}-\left| S_{N}\left( {\bf B}\right) \right| ^{2}}{J_{N}+\left|
S_{N}\left( {\bf B}\right) \right| \sin \left( \omega _{B}\cdot \,t-\alpha
_{B}\right) }
\]
For a static magnetic field ${\bf B}$ the voltage response function $\left\langle
V_{xy}\right\rangle $ measured between the nodes $x$ and $y$ of the circuit
{\it \ }is equal to the {\it dc}-part of the rapidly oscillating voltage
signal $V_{0}(t)$. All Josephson junctions in the {\it 1D }array
oscillate at the same frequency $\omega_{B}=2\pi \, \nu _{B}$, which is related to 
$\left\langle V_{xy}\right\rangle $ by: 
\begin{equation}
\frac{h}{2e}\nu _{B}=\left\langle
V_{0}\right\rangle =I_{c}R\sqrt{J_{N}^{2}-\left| S_{N}\left( {\bf B}\right)
\right| ^{2}}=\left\langle V_{xy}\right\rangle  \label{Einstein}
\end{equation}
%=\frac{\hbar }{2e}\omega _{B}
Note that the oscillation frequency $\nu _{B}$ of
such a local oscillator is even more sensitive to changes of strength or
orientation of the external magnetic field ${\bf B}$ than the structure
factor of the array itself, since $\left| S_{N}\left( {\bf B}\right) \right| 
$ enters Eq.(\ref{Einstein}) {\em quadratically}.

Consider, as a special case, an ordered array, consisting of $N-1$ {\em identical} 
{\it SQUID} loops, such that 
$\left\langle {\bf B}\,,\,{\bf a}_{n}\right\rangle \,=\Phi =\left\langle {\bf B}\,,\,{\bf a}_{L}\right\rangle$, 
and $I_{c,\,n}=I_{c}$ independent on the junction index $n$. Then the
structure factor $S_{N}\left( {\bf B}\right) \equiv $ $S_{N}^{\left( \Phi\right) }$ 
becomes a simple geometrical series: 
\begin{equation}
S_{N}^{\left( \Phi \right) }=
\frac{\sin \left( \pi \frac{\Phi }{\Phi _{0}} N\right) }
{N\sin \left( \pi \frac{\Phi }{\Phi _{0}}\right) }\exp \left[ \pi i \frac{\Phi }
{\Phi _{0}}\left( N-1\right) \right]
\end{equation}
In  Fig.(1b) one observes the usual narrowing proportional to $\frac{1}{N}$ of 
the width of the voltage response signal $\left\langle V_{xy}\right\rangle $ 
around its minima.
Note the periodicity property $\left| S_{N}^{\left( \Phi +\Phi
_{0}\right) }\right| =\left| S_{N}^{\left( \Phi \right) }\right| $ for all
$N\geq 2$. For $N=2$ Eq.(\ref{Einstein}) is the periodic voltage response
of a symmetric two junction {\it SQUID }in the
overdamped junction regime\cite{Likharev}.

A structure factor with a much longer period may be obtained in a parallel
junction array where the orientated area elements increase in size according
to a {\em linear} relation: 
\begin{equation}
{\bf a}_{m}=(2m-1)\,{\bf a}_{1}  \label{linear area law}
\end{equation}
For simplicity, identical junction parameters $R_{n}$, $C_{\,n}$ and $I_{c,\,n}$ 
are assumed. Then: 
\begin{equation}
S_{N}({\bf B})=\frac{1}{N}\sum_{n=0}^{N-1}\exp \left[ 2\pi i
\frac{\left\langle {\bf B}\,,\,{\bf a}_{1}\right\rangle }{\Phi _{0}}n^{2}\right]
\end{equation} \nolinebreak
The total area occupied by such a {\it Gaussian} array is $\left( N-1\right)
^{2}{\bf a}_{1}$, where ${\bf a}_{1}$ is the smallest area element, and 
${\bf a}_{N-1}=\left( 2N-3\right) {\bf a}_{1}$ is the largest area element.
Compare a {\it Gaussian} array, with $N-1$ area elements as described in 
Eq.(\ref{linear area law}), with a {\it periodic} array, consisting of $N_{P}-1$
identical {\it SQUID}-loops with size $\left| {\bf a}_{L}\right| $. For a useful 
comparison, both arrays should occupy the same total area: 
$\left( N-1\right) ^{2}{\bf a}_{1}=\left( N_{P}-1\right) {\bf a}_{L}$. Also
the largest area element in the {\it Gaussian} array should coincide with
the area element of an optimal single {\it SQUID}-loop, i.e. 
${\bf a}_{N-1}={\bf a}_{L}$. Both requirements together imply for $N_{P}\gg 1$ that the 
{\it Gaussian} array has the double number of junctions compared to a
corresponding periodic junction array: $N\simeq 2N_{P}$. To determine the
period of the Gaussian array consider a case where the flux threading the
area of the smallest element, ${\bf a}_{1}$, is equal to a rational multiple
of half a flux quantum: $\left\langle {\bf B}\,,\,{\bf a}_{1}\right\rangle
\,=\frac{M}{N}\frac{\Phi _{0}}{2}$. Then the largest area element in the
array, ${\bf a}_{N-1}$, is threaded by a flux 
$\Phi _{M}=\left( 1-\frac{3}{2N}\right) M\Phi _{0}$. In this case the structure factor $S_{N}({\bf B})
\equiv S_{N}^{\left( \Phi _{M}\right) }$ may be determined using a result
of C.F. Gauss\cite{Lindelöf}: 
\begin{equation}
S_{N}^{\left( \Phi _{M}\right) }=\frac{1}{N}\sum_{n=0}^{N-1}e^{\pi i\frac{M}{N}\,
n^{2}}=\frac{e^{i\frac{\pi }{4}}}{\sqrt{N\cdot M}}\sum_{n=0}^{M-1}e^{-\pi i\frac{N}{M}n^{2}}
\label{Gauss Sum}
\end{equation}
Note the periodicity $\left| S_{N}^{\left( \Phi _{M}+\Phi _{2N}\right)
}\right| =\left| S_{N}^{\left( \Phi _{M}\right) }\right| $, with period 
$\Phi _{2N}=\left( 2N-3\right) \Phi _{0}$. Remarkably, for $M=2$, and 
$N=N_{1}N_{2}$ being the product of two prime numbers $N_{1}$ and $N_{2}$,
there holds the factorization: 
\begin{equation}
S_{N}^{\left( \Phi _{2}\right) }=\left( -1\right) ^{\frac{\left(
N_{1}-1\right) \left( N_{2}-1\right) }{4}}S_{N_{1}}^{\left( \Phi _{2}\right)}
S_{N_{2}}^{\left( \Phi _{2}\right) }
\end{equation}
Apparently, such {\it Gaussian} junction arrays are governed by the laws of
number theory (quadratic residues\cite{Schröder}).

The long periodicity of the structure factor vs. flux $\Phi ^{\left( 1\right)}$ threading the
largest area element ${\bf a}_{L}$ of {\it Gaussian} junction arrays is also
visible in the calculated voltage response function $\left\langle
V_{xy}\right\rangle $, irrespective of the degree of the inductive coupling
represented by the parameter $\beta _{L}$. This is illustrated in Fig.(2).
Note the asymmetry of  $\left\langle
V_{xy}\right\rangle $  under $\Phi \rightarrow -\Phi $ (for a constant bias
current $I$) for finite inductive coupling. As far as disorder is
concerned, we also find that $\left\langle V_{xy}\right\rangle $ in {\it Gaussian} 
junction arrays is very responsive to adding small random
fluctuations to the size distribution of the area elements, so that 
$\left\langle V_{xy}\right\rangle $ becomes non periodic with a pronounced
antipeak only around $\Phi =0$. As $\beta _{L}$ increases, the difference 
$\max\left\langle V_{xy}\right\rangle -\min\limits\left\langle V_{xy}\right\rangle$ 
decreases, and the linewidth of the
global minimum $\left\langle V_{xy}\right\rangle $ around $\Phi =0$ ceases in this case
to scale proportional to $\frac{1}{N}$. Note all this applies in the
overdamped junction regime. Our results for weak damping will be published
elsewhere \cite{qif2}. 

We hope that an experimental verification of the predicted magnetic field
dependence of the voltage response function of {\it disordered} $1D$
parallel Josephson junction arrays will stimulate the development of new
types of robust superconducting quantum interferometers, which would allow
(for the first time) a technically rather simple precision measurement of 
{\it absolute} strength of external magnetic fields.
\begin{figure}
	\resizebox*{79 mm}{!}{\includegraphics{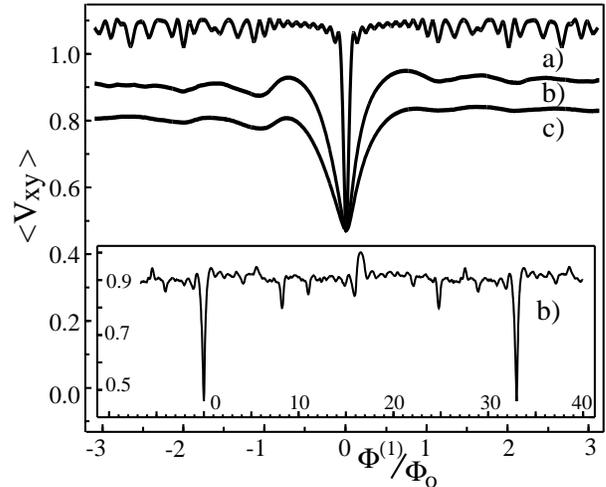}}
\caption{Voltage response $\left\langle V_{xy}\right\rangle$  in units of $I_c \, R$ vs. external flux 
$\Phi ^{\left( 1\right)}$ through largest area element ${\bf a}_{L}$ for a {\it Gaussian} array 
with $N=18$ (overdamped) junctions for  bias current $I=1.1\,N\,I_c$ and various inductive couplings: 
a) $\beta_{L}=0$, b) $\beta_{L}=0.3$ and c) $\beta_{L}=0.7.$}
\end{figure}
{\bf Acknowledgments}: We thank R.P. Huebener, R. Kleiner and T. Tr\"{a}uble
for useful discussions. Support by ''Forschungsschwerpunktprogramm des
Landes Baden-W\"{u}rttemberg'' is gratefully acknowledged.
%%%%%%%%%%%%%%%%%%%%%%%%%%%%%%%%%%%%%%%%%%%%%%%%%%%%%%%%%%%%%%%%%%%%%%%%%%%
% Bibliography                                                            %
%%%%%%%%%%%%%%%%%%%%%%%%%%%%%%%%%%%%%%%%%%%%%%%%%%%%%%%%%%%%%%%%%%%%%%%%%%%

\end{document}